\documentclass{aa}  
\usepackage{graphicx}
\usepackage{txfonts}

\usepackage{color, colortbl}
\usepackage[hidelinks]{hyperref}
\usepackage{xcolor}
\hypersetup{
    colorlinks = true, 
    linkcolor={red!50!black},
    citecolor={blue!50!black},
    urlcolor={blue!80!black}
}
\usepackage{amstext}

\begin{document} 

\definecolor{bubbles}{rgb}{0.91, 1.0, 1.0}
\definecolor{columbiablue}{rgb}{0.61, 0.87, 1.0}
\definecolor{cream}{rgb}{1.0, 0.99, 0.82}
\definecolor{lightblue}{rgb}{0.68, 0.85, 0.9}
\definecolor{lightcyan}{rgb}{0.88, 1.0, 1.0}
\definecolor{antiquewhite}{rgb}{0.98, 0.92, 0.84}
\definecolor{champagne}{rgb}{0.97, 0.91, 0.81}
\definecolor{electricblue}{rgb}{0.49, 0.98, 1.0}
%
        
%
   \title{Secondary eclipses of two brown dwarfs in the K2 fields:   
          detection by multiple dataset merging}
   \author{Geza Kovacs 
          \inst{1} 
	  \and 
          Behrooz Karamiqucham  
          \inst{2}
	  \and 
          J\'anos Zsargo
          \inst{3}
          }

   \institute{Konkoly Observatory, Research Center for Astronomy and Earth Sciences  
              of HUN-REN, MTA Center of Excellence,  
              Budapest, 1121 Konkoly Thege ut. 15-17, Hungary\\
              \email{kovacs@konkoly.hu}
	      \and
	      Department of Physics \& Astronomy, College of Charleston, 
	      Rita Hollings Science Center, 58 Coming Street, Charleston, SC 29421, USA
	      \and
	      Escuela Superior de F\'{i}sica y Matem\'aticas, Instituto Polit\'ecnico Nacional (IPN), 
	      Luis Enrique Erro S/N, San Pedro Zacatenco, 07738 Ciudad de M\'exico, Mexico
             }

   \date{Received 09-09-2025 / Accepted DD-MM-202?}


%
%
  \abstract
{By using various data sources for the stellar fluxes in overlapping 
campaign fields and employing full time series modeling, we report 
the detection of the secondary eclipses of two brown dwarfs (CWW~89Ab = 
EPIC~219388192b and HSHJ~430b = EPIC~211946007b). The detections 
yielded timings in agreement with the orbital elements derived 
from the earlier radial velocity measurements and eclipse depths 
of $70\pm 12$~ppm (CWW~89Ab) and $852\pm 123$~ppm (HSHJ~430b). 
While the high depth in the Kepler waveband for HSHJ~430b is in 
agreement with the assumption that the emitted flux comes mostly 
from the internal heat source and the absorbed stellar irradiation, 
the case of CWW~89Ab suggests very high albedo, because of the lack 
of sufficient thermal radiation in the Kepler waveband. Assuming 
completely reflective dayside hemisphere, without circulation, 
the maximum value of the eclipse depth due to the reflection of 
the stellar light is $56$~ppm. By making the extreme assumption that 
the true eclipse depth is $3\sigma$ less than the observed depth, 
the minimum geometric albedo becomes $\sim 0.6$.  
}


\titlerunning{Secondary eclipses of two brown dwarfs}
\authorrunning{Kovacs et al.}

   \maketitle
%
%
%
\section{Introduction}
\label{sect:intro}
Being at the borderline of planets and `classical' stars, brown 
dwarfs (BDs) play a specific role in our understanding both the 
evolution of very low mass stars and the atmospheres of hot giant 
planets. Of particular interest are those BDs that are members of 
binary systems, and, especially, if they are also eclipsing systems. 
The apparent rarity of BDs within $5$~AU of their main sequence 
hosts has led to the problem of ``brown dwarf desert'', as derived 
from early radial velocity surveys \citep[][]{marcy2000, halbwachs2000}. 
Although there has been a very spectacular increase in the number 
of transiting BD systems during the past several years, dominantly 
aided by the Transiting Exoplanets Survey Satellite 
\citep[TESS, see][]{ricker2015}, the BD desert phenomenon seems to survive 
\citep[albeit with a lower contrast, see][]{carmichael2019, stevenson2023, vowell2025}.  

Secondary eclipse observations play a significant role in extrasolar 
planetary atmosphere science and it has similar importance in the 
study of BDs. Due to their internal heat source, under favorable 
conditions, in general, BDs are better candidates for 
secondary eclipse studies than close-in exoplanets, as the observed 
drop in the flux during the eclipse event is less dependent on the 
unknown (and  usually low) geometric albedo (while this statement 
is true in the infrared, in the visible the situation might be more 
involved). 

In spite of their favorable position for occultation detections, the 
number of the published successful detections is relatively low. 
Using the list of \cite{barkaoui2025} and extending it by a few 
additional systems, within the classical BD mass limit of 
$\sim 80$~M$_{\rm J}$, there are $55$ objects. From these, $8$ systems 
have at least one successful secondary eclipse detection, some of 
them in multiple wavebands 
\citep[e.g., LHS~6343=KOI-959, see][]{frost2024}. The purpose of 
this paper is to add to this list two new detections 
in the Kepler passband and investigate the possible constraints 
that these detections might pose on the orbit and, in particular,
on the atmospheric properties (i.e., on the albedo). 

The two systems to be investigated are members of nearby, 
deeply-studied open clusters, and therefore, we have well-defined 
ages, distances and constraints on their chemical compositions. 
Age, in particular, is a prime factor in determining the basic 
physical paramters -- mass, radius and effective temperature 
\citep[see, e.g.,][]{baraffe2003}. Consequently, the two systems 
discussed in this paper are among the most significant binaries 
with BD secondaries.   

The number of transiting BD systems with proven cluster 
membership is rather low. According to \cite{acton2021}, there 
are only three BD systems in clusters (two of these are discussed 
in this paper, and a third one, RIK-72b (EPIC 205207894b), is in 
the Upper Scorpius OB association \citep{david2019}.\footnote{By 
examining the list of \cite{barkaoui2025}, we also found that 
HIP~33609 is a member of the MELANGE-6 association \citep{vowell2023}.} 
This low number of transiting BDs in clusters is in contrast with 
the high number of BDs observed through imaging surveys of clusters. 

The first system is CWW~89A (EPIC~219388192), a member of  
Ruprecht~147, a 2.67~Gyr old cluster. This is a triple system, with 
a solar-type main sequence host, and a distant outer companion 
\citep[a likely M dwarf at $\sim 25$~AU, see][]{curtis2016, nowak2017, beatty2018}.  
Interestingly, the observed eclipse depths in two Spitzer wavebands 
suggest a much higher internal temperature than current evolutionary 
models predict \citep[see][]{beatty2018}. Therefore, it is important 
to examine if this discrepancy has any trace also at the shorter 
passband of Kepler.  
 
The second system is HSHJ~430\footnote{We use the name suggested by 
SIMBAD (\url{https://simbad.cds.unistra.fr/simbad/}) for the apparently  
non-traditional earlier naming of AD~3116, that is simply a record 
number assigned by the VizieR team.} (EPIC~211946007), a member of 
the open cluster Praesepe. It consists of an M dwarf and a 
BD \citep{gillen2017}. Here we present the first occultation 
detection for this system. 

An integral part of the methodology employed in this paper is the 
utilization of the unique data availability for the K2 mission. 
The different approaches employed by various research groups to 
derive stellar fluxes enable us to search for shallow signals with 
greater success even in the case of single field target occupancy. 
Without the various aperture photometry sources and overlapping 
campaign fields (for EPIC~211946007) we could not detect the 
signals reported here (or, at least, their significance would have 
been far lower).

%
%
\section{Method}
\label{sect:method}
There are three major steps in the analysis of each object. First, 
we treat the data on an author-by-author (hereafter source-by-source) 
and (campaign) field-by-field basis -- see Sect.~\ref{sect:data} 
for the specification of the individual datasets. Then, once all 
the time series and statistical information is saved, these are 
fed into the data merging code that also performs the signal search on 
the averaged time series and statistics. Finally, in the third step, 
the single parameter (namely, $\delta_2$, the secondary eclipse 
depth in the Kepler passband Kp) is compared with the theoretical 
models and its consistency with other pieces of information is checked. 
We discuss the first two steps in the following subsections. More 
detailed discussion concerning the theoretical interpretation is 
postponed to Sect.~\ref{sect:models}.

%
%
\subsection{Single time series analysis}
\label{sect:method1}
The data treatment and modeling of the single source, single 
field time series is akin to those employed in our earlier papers 
\citep{kovacs2020,kovacs2025}, extended by the specific task of 
searching for eclipse features with the same orbital period and 
eclipse shape as the already known transit signal. The observed 
(Simple APerture -- SAP) flux $X(t)$ is approximated by the 
following additive signal model: 
%
%
\begin{eqnarray}
\label{ts_model}
X(t) = FOUR(t) + SYS(t) + OOTV(t) + TR(t) + EC(t)\hspace{2mm}. 
\end{eqnarray}
Here $FOUR(t)$ is the Fourier representation of the stellar 
variability. This implies usually a high-order 
(i.e., $m_{\rm FOUR}\sim 90$) Fourier fit, with the frequencies 
of $1/T, 2/T, ...,m_{\rm FOUR}/T$, where $T$ stands for the full 
length of the given K2 observation campaign. We note that this 
is a general Fourier decomposition, and does not require the 
knowledge of the nature of the variability. It produces a 
reliable representation of non-stationarity, often present 
in stellar activity induced flux changes. 

The corrections due to instrumental systematics are represented 
by $SYS(t)$. This involves both common features in many stars 
in the same field \citep[see, e.g.,][]{smith2012} and those 
specific to the target \citep[i.e., $x, y$ pixel position- and $bg$ 
background-dependent flux changes -- see][]{bakos2010}. 
The common systematics are derived from randomly selected stars, 
covering each CCD module roughly in equal number. These are 
passed through various non-variability criteria by ending up 
with some $500$ non-variable stars for the full Kepler field of 
view. Next, an orthogonal set is built up from these time series 
with the aid of a Principal Component Analysis (PCA) routine, 
that yields also the order of `strength of commonality' via the 
eigenvalues of the PCA vectors. After a detailed experimentation, 
we found that using the first $100$ PCA eigenvectors is an 
optimum choice to reach efficient eclipse signal detection. 

The out-of-transit variation $OOTV(t)$ is represented by a 
second-order Fourier sum ($m_{\rm OOTV}=2$), allowing a rough modeling 
any phase and tidal variation with the orbital frequency and 
its first harmonic. It is important to note that although we 
introduce these orbital motion related variations, in the case 
of the two objects discussed in this paper, they will be largely 
eliminated by the high-order Fourier fit needed to tackle stellar 
variability, due to spots. Therefore, the derived $OOTV(t)$ will 
be considered only as an additional correction function to eliminate 
non-eclipse-like variations related to the orbital motion.   

The transit $TR(t)$ is represented by a trapezoid with an 
U-shaped bottom part, similar as described by \cite{kovacs2020}. 
The eclipse shape $EC(t)$ is a flat-bottomed trapezoid with 
the same eclipse and ingress durations (respectively, {\em t14} 
and {\em t12}) as those of the transit.

While the determination of the correction functions represented 
by the linear combinations of known functions is straightforward 
via weighted least squares, that of \{$TR$\} and \{$EC$\} requires 
multistep methods. For \{$TR$\} we start from the roughly known 
transit parameters and employ a simple Monte Carlo (MC) method 
to sample of the neighborhoods of those parameters. The period 
is fixed in this process, and taken as the average as given by 
our earlier analyses using the datasets dealt with in the present 
analysis. 

In the case of the eclipse function \{$EC$\}, the situation 
is more complicated, because of the unknown eclipse phase. 
For this reason, first we need to compute the residual 
time series \{$Y$\}, left after the subtraction of all known 
signal constituents, namely, 
$Y(t)=X(t)-FOUR(t)-SYS(t)-OOTV(t)-TR(t)$. Then, using the 
folded version of $Y(t)$ (phased by the orbital period), we 
scan this function by shifting the trial eclipse center 
throughout the full orbital phase\footnote{Although the 
U-bottomed transit approximation works well in most cases, 
for the sake of the more sensitive shallow secondary eclipse, 
we mask out the transit phase by substituting \{Y\} with a 
Gaussian noise of the same standard deviation as the rest 
of \{Y\}.} and fit the eclipse depth with the aid of weighted 
least squares. It is important to note that the determination 
of the weights is not part of the scanning process, but they 
are fixed and derived after each full iteration sequence, 
involving all known signal components. By considering the 
phase dependence of the fitted eclipse depth, we arrive to 
the Secondary Eclipse Search (SES) statistics, the basic 
information needed to make decision about the significance 
of the detection. 

%
%
\begin{figure}[t]
\centering
\includegraphics[width=0.45\textwidth]{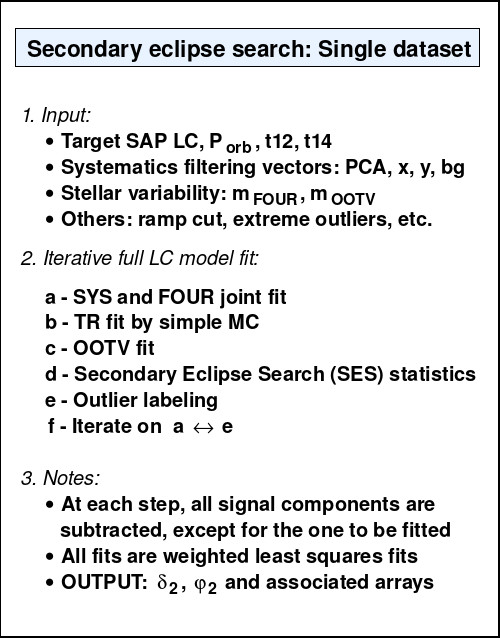}
\caption{Brief summary of the parameters and steps playing role in 
         the analysis of the single datasets. See text for additional 
	 details.} 
\label{single_chart}
\end{figure}

The decomposition shown by Eq.~\ref{ts_model} is performed 
through an iterative scheme, during which we fit each entity 
separately\footnote{Except for \{SYS\} and \{FOUR\}, that 
are fitted jointly.} by subtracting the other, already known 
(or estimated) functions from the input signal \{$X$\}. 
The iteration converges within $\sim 10$ iterations. The 
resulting time series is labeled as `reconstructed', whereas 
the one that is the result of only one iteration cycle is 
referred to as `non-reconstructed'. Naturally, full reconstruction 
results in a better approximation of the transit 
\citep[or other signal constituents -- see ][]{kovacs2005}. 
Nevertheless, in the case of shallow signals, the component might 
be falsely identified and the reconstruction may amplify that 
misidentified noise components. Therefore, in faint signal search 
the examination of the statistics related to the non-reconstructed 
time series is also important.  

For a quick summary, the procedures discussed in this subsection 
are shown in Fig.~\ref{single_chart}. Not shown in the figure, 
and not mentioned so far is the role of background stars. They   
may change the depth of the eclipse, and therefore, affect the 
success of the signal search on the averaged data and introduce 
additional error in the derived parameters. Therefore, a part of 
the analysis and preparation of the data input for the merging 
phase, is the determination of the hidden flux corrections for 
each dataset. This has been done by a trial and error method, 
whereby we used the thought to be most reliable transit depth 
determination and scaled all time series to yield the transit 
depth chosen. The size of the corrections were around a few 
percentages, but, due to the high-level of overall crowdedness 
of field C07, for EPIC 219388192, for some of the sources, 
the corrections were around $10$\% (see Appendix~\ref{app_A} 
for additional details).

%
%
\subsection{Multiple time series analysis}
\label{sect:method2}
Here the time series and SES statistics produced by the analysis 
of the target-, source- and field-specified data (as described in 
Sect.~\ref{sect:method1}) are fed into a code that combines these 
data for a given target with the goal of increasing the signal-to-noise 
ratio (SNR) of the unknown signal component. Due to their different 
meanings, time series and SES statistics are averaged differently. 
This allows us to compare the two results, serving as a sanity 
check for the final conclusion. 

When averaging the SES statistics corresponding to the individual 
datasets, we focus on the possible signatures of a signal showing 
up in SES. The signal is exhibited as a dominant dip in this function. 
Therefore, it is meaningful to characterize the relevance of this 
particular SES, as the ratio between the largest dip ($\delta_2$) 
and the scatter (i.e., $\sigma(\rm{SES})$, the standard deviation)  
of SES: 
%
%
\begin{eqnarray}
\label{ses_snr}
{\rm SNR} & = & {|\ \delta_2-\langle \rm{SES} \rangle\ | \over \sigma(\rm{SES})}  
\hspace{2mm},  
\end{eqnarray}
where $\langle\rangle$ denotes the average of SES. In computing 
$\langle \rm{SES} \rangle$ and $\sigma(\rm{SES})$, we exclude the phase 
range of $\pm t14/P_{\rm orb}$ around $\delta_2$, to avoid downward 
biasing of SNR. The eclipse depth $\delta_2$ is searched in a 
limited phase range (i.e., between $0.35 < \varphi < 0.65$, a relatively 
close neighborhood of the expected phase of $0.50$, in the case of 
zero eccentricity). Although this limitation of the search for 
$\delta_2$ introduces  some bias in the final result (i.e., a preference 
toward low-eccentricity solutions), if the main dip in the restricted 
region is small, then the corresponding weight on the related 
SES will also be small. Therefore, the bias is expected to be 
small. The standard phase range above can be changed if there 
is a sufficient evidence that the main dip is outside this range. 
In the two targets discussed in this paper, there were no such 
evidences.  

%
%
\begin{figure}[t]
\centering
\includegraphics[width=0.45\textwidth]{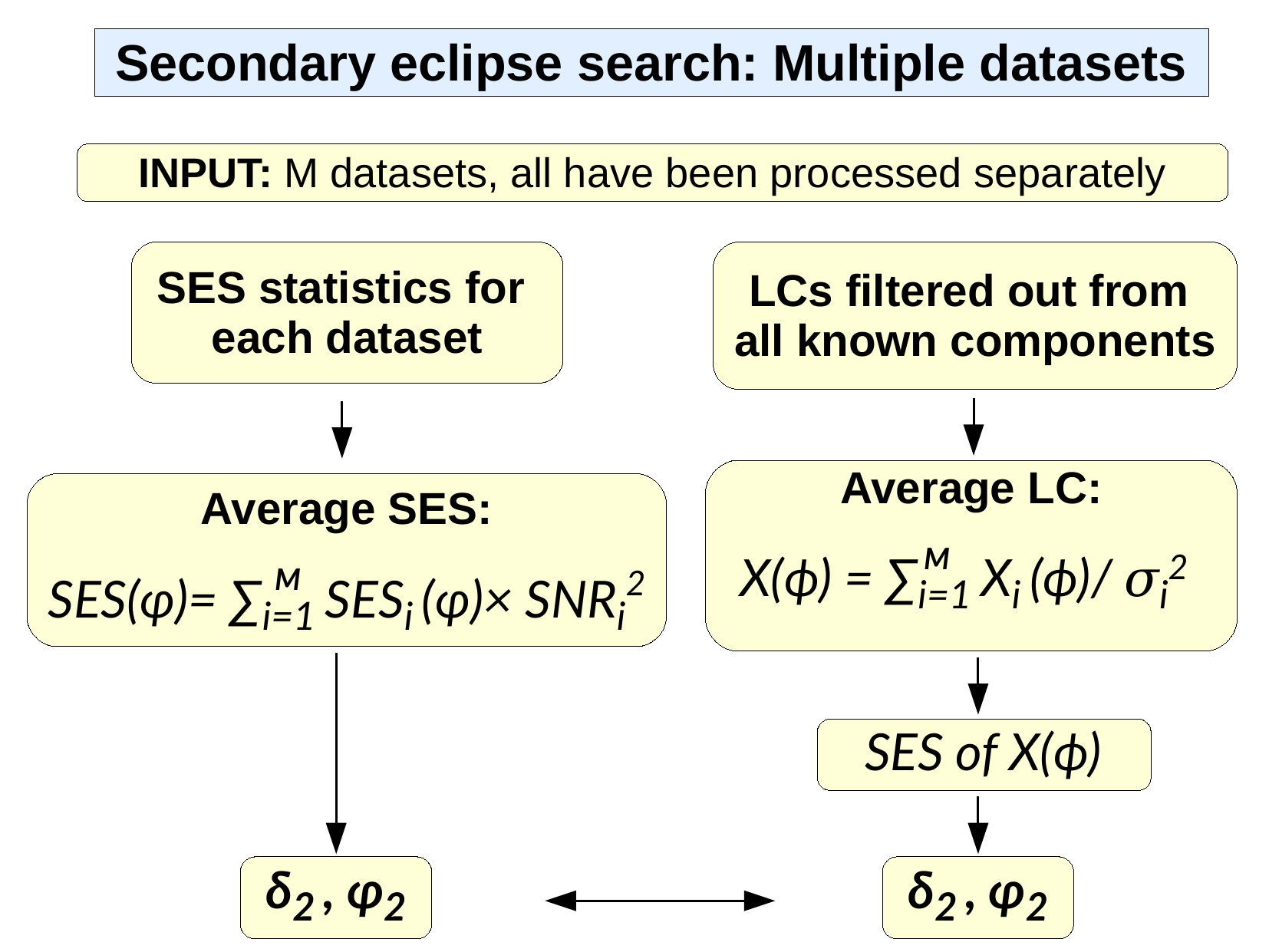}
\caption{Two types of approach in searching for secondary eclipse 
         by using multiple datasets (already filtered out from all 
	 other components -- see Sect.~\ref{sect:method1} and 
	 Fig.~\ref{single_chart}). 
	 {\em Left:} averaging SES statistics of the individual 
	 datasets by their SNR. 
	 {\em Right:} averaging individual LCs by their inverse 
	 variance, and then compute SES of the average LC. The sum 
	 of the weights are normalized to unity in both cases. 
	 See text for more.} 
\label{multiple_chart}
\end{figure}

The left part of Fig.~\ref{multiple_chart} displays the 
simple process of SES weighting. The output parameters are 
$\delta_2$, the depth of the eclipse, and its phase, $\varphi_2$ 
(relative to the phase of the transit, i.e., 
$\varphi_2 = \phi_{\rm EC} - \phi_{rm TR}$).    
One may also average the individual LCs, and then compute the SES 
of the average LC (see the right part of Fig.~\ref{multiple_chart}).
While in the case of SES averaging we have arrays sampled at the 
same phase values, the LCs -- even if they are folded -- are not 
sampled in the same way. Therefore, we need not only to fold them, 
but also bin them on the same phase grid. We found that $1500$ 
bins throughout the full orbital phase yield fine eclipse resolution 
and also, relatively few empty bins. 

A natural way of averaging the above binned/folded LCs is to 
employ inverse variance weighting. The average LC will not carry 
any information about our preference of dip selection, like in the 
case of SES averaging. Because the unknown signal is weak, the 
variances are expected to be good approximations of the true variances. 
The average LC then undergoes of the standard SES analysis and another 
estimation is obtained for the two parameters ($\delta_2$ and $\varphi_2$) 
of the secondary eclipse. 

Although the different data sources yield somewhat different LCs, 
the associated stochastic components (naturally) are not independent 
(i.e., all sources have some common pixels while evaluating the 
total flux). Therefore, the noise averaging will not follow the 
classical $\sim1/\sqrt n$ law, but it will decrease with a 
lower pace as a function of the number of data sources. We found 
that the noise of the averaged LC is some $40$--$50$\% higher than 
expected from averaging independent random numbers. Even so, 
this milder improvement in the data quality is already quite 
significant, when the signal is at the verge of detectability for 
the single datasets.    

%
%
\section{Datasets}
\label{sect:data}
As described in \cite{kovacs2025}, for K2, there are at least four 
easy to access datasets. All these sets were produced by using 
different methods in generating the raw time series from the 
long-cadence SAP fluxes. 
%
%
\begin{table}[h]
\begin{flushleft}
\caption{Basic properties of the input time series}
\label{input_ts}
\scalebox{1.0}{
\begin{tabular}{ccccrc}
\hline
Field & Source &   N   &    RMS    & $N_{\rm clip}$ & $T$\\
\hline\hline
\multicolumn{6}{c}{EPIC 219388192}\\
\hline
  C07 &  LUG   & 3846  & 0.000117  &  128  & 83 \\
  C07 &  KEP   & 3429  & 0.000157  &   98  & 83 \\
  C07 &  PET   & 3475  & 0.000187  &   89  & 83 \\
  C07 &  VAN   & 3721  & 0.000234  &  108  & 83 \\
\hline
\multicolumn{6}{c}{EPIC 211946007}\\
\hline
  C05 &  LUG   & 3663  & 0.002294  &  112  & 74 \\
  C05 &  KEP   & 3357  & 0.002639  &   73  & 74 \\
  C05 &  PET   & 3620  & 0.003454  &  106  & 74 \\
  C05 &  VAN   & 3402  & 0.002987  &   83  & 74 \\
  C16 &  LUG   & 3886  & 0.002406  &  157  & 80 \\
  C16 &  KEP   & 3563  & 0.002796  &  109  & 80 \\
  C16 &  PET   & 3703  & 0.003377  &  131  & 80 \\
  C16 &  VAN   & 3244  & 0.003342  &   90  & 80 \\
  C18 &  LUG   & 2472  & 0.002808  &   81  & 51 \\
  C18 &  KEP   & 2364  & 0.002371  &   72  & 51 \\
  C18 &  VAN   & 2265  & 0.003424  &   77  & 51 \\
\hline
\end{tabular}}
\vskip 4 pt
\begin{minipage}{0.49\textwidth}
{\bf Notes:}{\small\
Abbreviations used in the source names are described in the text. 
${\rm N}=$ total number if input data points; RMS$=$ root mean 
square of the residual flux after subtraction of the full time series 
model from the input data; the RMS values have been corrected for 
data clipping; $N_{\rm clip}=$ number of $3.3\sigma$-clipped items. 
The campaign duration ($T$) is given in days.}\\ 
\end{minipage}
\end{flushleft}
\end{table}
We employ these raw LCs from the following sources. 
Starting from the output of the pipeline of the mission, we 
downloaded the ASCII files from the IPAC exoplanet 
site\footnote {\url{https://exoplanetarchive.ipac.caltech.edu/}}. 
These data are labeled by KEP. We note that Space Telescope Science 
Institute's Mikulski Archive for Space Telescopes (MAST) 
site\footnote{\url{https://archive.stsci.edu/}} also hosts  
nearly the same data, but in our experience they contain far 
larger number of outliers, making the analysis more cumbersome 
and, ultimately, leaving more noise in the LCs. The next set 
by \cite{petigura2015} (hereafter PET) comes from the earlier 
version of the NASA ExoFop site from the main Exoplanet Archive 
as cited earlier. The other two sets by \cite{vanderburg2014} (VAN) 
and \cite{luger2016,luger2018} (LUG) have been downloaded from the 
corresponding MAST sites.   

The list of datasets used in the time series analysis is given in 
Table~\ref{input_ts}. All parameters refer to the residuals obtained 
from the non-reconstructed LCs (with reconstruction the parameters 
do not change at any significant degree). For C18 we have only three 
sources, because of the incompleteness of the PET data for this 
campaign at the moment of download attempt at the IPAC ExoFop site. 
The two objects have drastically different noise levels, due to the  
large difference in their brightness ($16.57$~mag vs $12.34$~mag in 
the Kepler filter for EPIC 211946007 and EPIC 219388192, respectively). 

%
%
\section{EPIC 219388192b}
\label{sect:cww89}
Here we present the search diagnostics both for the reconstructed and 
for the non-reconstructed time series. Likewise, the effect of multiple 
data sources is demonstrated both by averaging of the individual SES 
statistics and by running the same SES routine on the averaged LC. 
As discussed in Sect.~\ref{sect:method2}, the latter comparison 
yields an additional test on the reliability of the detection, due 
to the difference between the ways how the averages were derived. 

%
%
\begin{figure}[t]
\centering
\includegraphics[width=0.40\textwidth]{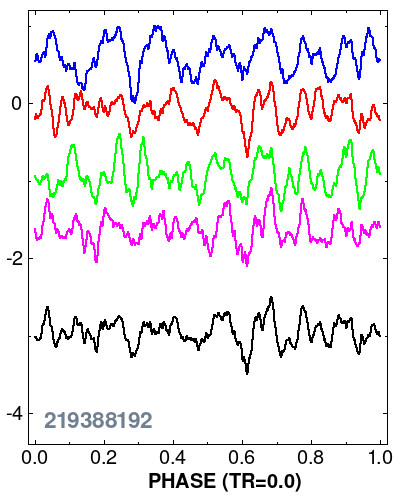}
\caption{SES statistics for the non-reconstructed LCs of EPIC 219388192. 
         From top to downward, plotted are SES statistics derived on the 
	 data sources of LUG, KEP, PET and VAN. The averaged SES is shown 
	 by black at the bottom. The phase scan is made from the transit 
	 center (phase zero) throughout the full orbital phase. All curves 
	 are plotted on the same (but arbitrary) scale and shifted properly 
	 for good visibility.} 
\label{219388192_01_dip_stat}
\end{figure}

The analysis of the LCs of the individual sources were optimized 
for more efficient signal detection both by examining the result 
for the real data and by the recoverability of injected signals. 
These tests have led to slightly different parameter settings 
for the individual sources. For example, it turned out that 
for all sources, except for that of LUG, our method yields better 
LCs than those employed by the sources. As a result, for LUG only, 
we used the published, systematics-filtered LCs, instead of those 
derived by our code from LUG's SAP time series. We followed this 
routine also for EPIC 211946007. 

No pixel position induced flux corrections were needed in 
any of the datasets for EPIC 219388192, but PET became 
substantially less noisy by using time-dependent background 
correction. For LUG, we had to omit the first 50 most outlying 
points and cut the LC for the first three days.   

\begin{figure}[t]
\centering
\includegraphics[width=0.40\textwidth]{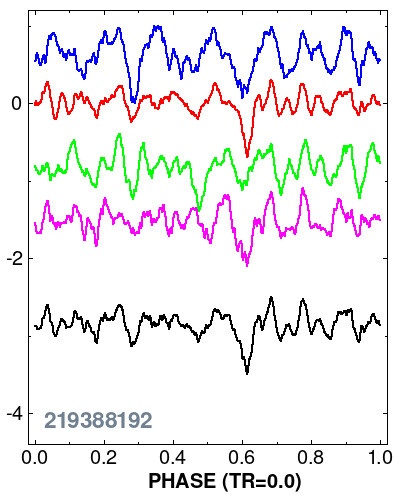}
\caption{As in Fig.~\ref{219388192_01_dip_stat}, but for the reconstructed 
         LCs.} 
\label{219388192_11_dip_stat}
\end{figure}

We searched for eclipse signals in the averaged LCs derived 
both from the non-reconstructed and from the reconstructed LCs 
of the four sources. Figure~\ref{219388192_01_dip_stat} shows the 
individual and the averaged SES statistics derived from 
the non-reconstructed LCs. The averaged SES reaffirms 
the presence of a signal suspected, in particular, in the KEP 
dataset. This finding is further strengthened by the inspection 
of the same diagram, using SES statistics derived from the 
reconstructed LCs (that use full signal models, including the 
updated eclipse signal during the reconstruction process). 
Figure~\ref{219388192_11_dip_stat} shows the corresponding 
SES statistics. In agreement with our expectation, the signal 
shows up even more clearly in the averaged SES. 

%
%
%
%

%
%
\begin{figure}[h]
\centering
\includegraphics[width=0.40\textwidth]{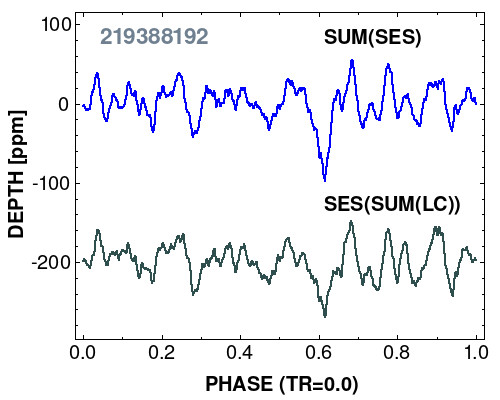}
\caption{Comparison of the final SES statistics obtained from averaging 
         the individual SES statistics (top) with the one obtained 
	 from the averaged LC (bottom). As for the earlier, similar 
	 plots, we employed arbitrary vertical shifts for better 
	 visibility. The results shown are based on reconstructed LCs.} 
\label{219388192_11_ses_compare}
\end{figure}

As discussed in Sect.~\ref{sect:method2}, an important part of the signal 
confirmation is finding the same signal on the basis of the SES analysis 
of the averaged LCs. Showing only the final SES statistics, 
Fig.~\ref{219388192_11_ses_compare} exhibits the presence of the same 
signal, independently of the method used. The lower SNR of the SES based 
on the averaged LC is not surprising, since in employing this method, 
LCs with higher scatter are penalized (no information on a possible 
underlying signal is used), whereas in the case of SES averaging, 
the low-SNR SES statistics are weighted lower (suspected signal 
information is used). 

%
%
\begin{figure}[h]
\centering
\includegraphics[width=0.40\textwidth]{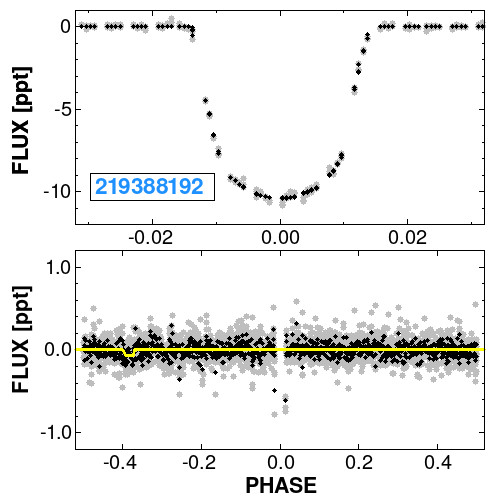}
\caption{{\em Upper panel:} Binned transit for the merged, reconstructed 
         LCs (gray), their U-bottomed trapezoid fits (black, 
	 binned and averaged in the same way as the data). 
	 {\em Lower panel:} Gray and black points as above, the yellow 
	 line shows the best-fitting flat-bottomed trapezoid for the 
	 secondary eclipse. The stroboscopic affect (exhibiting as small 
	 gaps) is not visible in the lower panel, because of the extended  
	 horizontal scale.} 
\label{219388192_11_aver_tr}
\end{figure}

The effect of averaging can also be exhibited by the LCs. Here we 
should use folded and binned LCs (see Sect.~\ref{sect:method2}). 
Figure~\ref{219388192_11_aver_tr} shows both the neighborhood of 
the transit and the close proximity of the OOT level. The nearly 
zero scatter among the U-bottomed trapezoid fits (upper panel, 
black dots), shows the effectiveness of the hidden flux correction 
applied to the data of the individual sources 
(see Appendix~\ref{app_A} for details). Without this 
correction, the transit depths scattered about $5-10$\%. As 
expected, the scatter of the average LC is smaller than those of 
the separate source data (black vs gray points in the lower panel). 
However, because the individual time series are not independent, 
the RMS is greater than the statistical limit ($70$~ppm vs $50$~ppm).  

Summary of the results obtained on the averaged SES statistics (method A) 
and those derived on the averaged LCs (method B) are given in 
Table~\ref{sec_219388192}. Because of the more involved nature of the 
eclipse depth error for method A, we give errors only for method B. 
Examining the derived parameters, the following trends can be 
observed. As expected, the reconstructed depths are larger, than the 
non-reconstructed depths, SNR-weighting for the SES statistics yields 
averaged SES with higher SNR than the one derived from the inverse 
variance weighted LCs. Preferring the complete signal modeling, but 
avoiding the chance of biased dip selection when method A is used, we 
settle on the result obtained by method B on the reconstructed LCs.   

%
%
\begin{table}[h]
\begin{flushleft}
\caption{Secondary eclipse parameters for EPIC 219388192}
\label{sec_219388192}
\scalebox{1.0}{
\begin{tabular}{ccccc}
\hline
Method & $\delta_2$  &  $e({\delta_2})$  &  ${\varphi_2}$  &  SNR$_{\rm SES}$ \\
\hline\hline
\multicolumn{5}{c}{Non-reconstructed}\\
\hline
  A &  72   & --  & 0.616  & 3.7 \\
  B &  50   & 12  & 0.616  & 2.6 \\
\hline
\multicolumn{5}{c}{Reconstructed}\\
\hline
  A &  97   & --  & 0.616  & 5.4 \\
\rowcolor{lightgray}
  B &  70   & 12  & 0.616  & 3.8 \\
\hline
  RV &  --   & --  & 0.617  & -- \\
\hline
\end{tabular}}
\vskip 4 pt
\begin{minipage}{0.49\textwidth}
{\bf Notes:}{\small\
Method A: final SES is the weighted sum of the SES statistics on 
the individual datasets. Method B: final SES is the SES statistics 
on the weighted sum of the individual folded and binned LCs. Errors 
for $\delta_2$ are the standard deviations of the means of the 
residuals within the eclipse. Eclipse depths and their errors are 
given in ppm. The preferred solution is shaded. The last row gives the 
phase of the secondary eclipse derived from the radial velocity data.}\\ 
\end{minipage}
\end{flushleft}
\end{table}

We see that all eclipse phases are the same for the various SES 
statistics, derived from the average LCs or individual SES statistics. 
Because the system has accurately measured radial velocity curves 
and subsequent analyses, it is possible to compare the above eclipse 
phase with those derivable from the spectroscopic orbital elements. 
Following \cite{winn2010} (his Eq.~33) and using the orbital elements 
$\omega$ and $e$ by \citep{carmichael2019}, we obtain\footnote{Because 
of its tiny effect ($10^{-4}$ in orbital phase units) we omitted the 
light-time effect for both systems discussed in this paper.} 
$\varphi_2 = 0.617\pm 0.002$. This is an excellent agreement, and 
lends an independent support of the reliability of the detected signal.

%
%
\section{EPIC 211946007b}
\label{sect:hshj430}
There are pros and cons for the detectability of the secondary eclipse 
of EPIC 211946007. First of all, the target is faint, leading to 
some $20$ times higher noise than in the case of EPIC 219388192. 
In addition, the host star is an M dwarf, with low flux level and, 
therefore, the reflected light is expected to be low. However, 
being member of the Praesepe cluster, the system is young, and 
therefore, the internal heat of the BD is expected to be high. 
Furthermore, the orbital period is only $1.983$~days, yielding 
favorable irradiation in spite of the cool M dwarf host. And last, 
but not least, the star is in the overlapping region of three 
K2 fields, allowing to suppress the high noise by the substantial 
number of data sources, covering three independent visits of the 
target.  

\begin{figure}[t]
\centering
\includegraphics[width=0.40\textwidth]{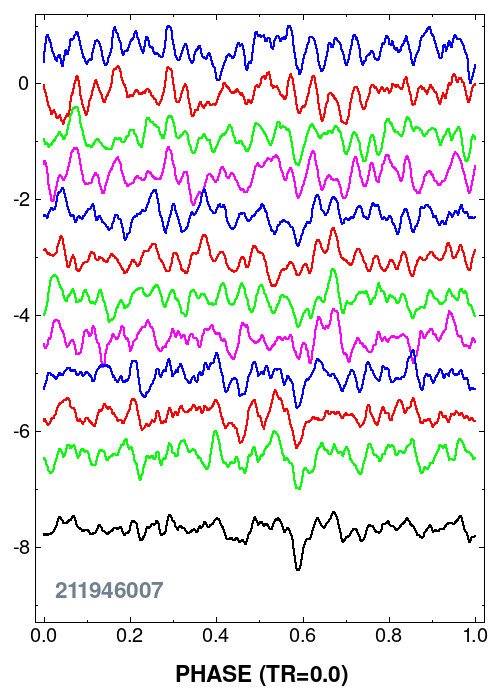}
\caption{SES statistics for the reconstructed LCs of EPIC 211946007. The 
         structure of the figure is similar to that of 
	 Fig.~\ref{219388192_01_dip_stat}. 
         From top to downward, we show the SES statistics derived on the 
	 data sources of LUG, KEP, PET and VAN. This pattern is repeated 
	 (from top to downward) for fields C05, C16 and C18 (for the latter, 
	 we have only LUG, KEP and VAN). Vertical shifting and scaling 
	 are applied as in Fig.~\ref{219388192_01_dip_stat}.} 
\label{211946007_11_dip_stat}
\end{figure}

Because of the great similarity of the non-reconstructed and 
reconstructed SES statistics (see also EPIC 219388192), 
we show only the relevant plots obtained from the reconstructed LCs. 
The SES statistics derived from the reconstructed LCs are shown in 
Fig.~\ref{211946007_11_dip_stat}. It is clear that the signal 
is barely visible in most of individual SES statistics. Due to the 
large number of data sources, and, importantly, to the field overlaps, 
the averaged statistics clearly exhibits the presence of a signal 
(more significantly, than in the case of EPIC 219388192).    

\begin{figure}[h]
\centering
\includegraphics[width=0.40\textwidth]{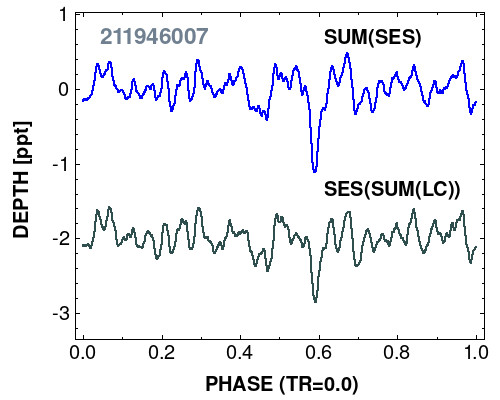}
\caption{Comparison of the final SES statistics obtained from 
         averaging of the individual SES statistics (top) with the 
	 one obtained from the averaged LC (bottom). The results 
	 shown are based on reconstructed LCs.} 
\label{211946007_11_ses_compare}
\end{figure}

\begin{figure}[h]
\centering
\includegraphics[width=0.40\textwidth]{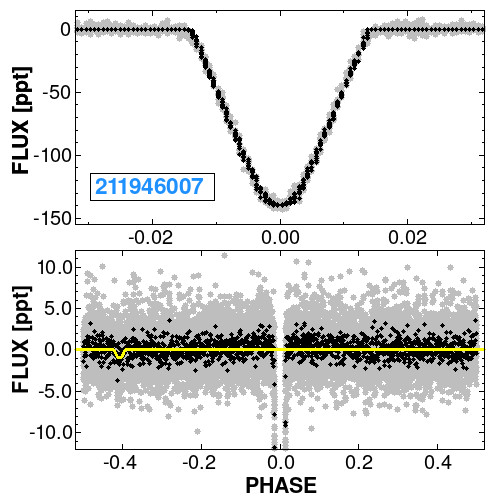}
\caption{{\em Upper panel:} Binned transit for the merged, reconstructed 
         LCs (gray), their U-bottomed trapezoid fits (black, 
	 averaged in the same way as the data). 
	 {\em Lower panel:} Gray and black points as above, the yellow 
	 line shows the best-fitting flat-bottomed trapezoid for 
	 the secondary eclipse.} 
\label{211946007_11_aver_tr}
\end{figure}

Testing the detection by using LC averaging first, and SES analysis 
after, on the average LC, confirms the detection 
(see Fig.~\ref{211946007_11_ses_compare}). 
The significant improvement on the LC quality due to averaging is 
exhibited in Fig.~\ref{211946007_11_aver_tr}.  

%
%
%
%

%
%
%
%

%
%
%

%
%
\begin{table}[h]
\begin{flushleft}
\caption{Secondary eclipse parameters for EPIC 211946007}
\label{sec_211946007}
\scalebox{1.0}{
\begin{tabular}{crccc}
\hline
Method & $\delta_2$  &  $e({\delta_2})$  &  ${\varphi_2}$  &  SNR$_{\rm SES}$ \\
\hline\hline
\multicolumn{5}{c}{Non-reconstructed}\\
\hline
  A &  918   &  --  & 0.591  & 5.3 \\
  B &  678   & 120  & 0.592  & 3.9 \\
\hline
\multicolumn{5}{c}{Reconstructed}\\
\hline
  A &  1110  & --  & 0.590  & 6.5 \\
\rowcolor{lightgray}
  B &   852  & 123 & 0.592  & 5.0 \\
\hline
  RV &  --   & --  & 0.592  & -- \\
\hline
\end{tabular}}
\vskip 4 pt
\begin{minipage}{0.49\textwidth}
{\bf Notes:}{\small\
Notation is the same as in Table~\ref{sec_219388192}}\\ 
\end{minipage}
\end{flushleft}
\end{table}

The resulting secondary eclipse parameters are shown in 
Table~\ref{sec_211946007}. Again, by using the orbital parameters 
of $\omega=5^{\circ} \pm 20^{\circ}$ and $e=0.146\pm 0.025$ from \cite{gillen2017} 
we arrive to $\varphi_2=0.593\pm 0.019$. Although the error is 
larger than for EPIC 219388192, the actual RV measurement is in 
nice agreement with the value derived from the independent 
photometric data.

%
%
\section{Comparison with evolution and atmosphere models}
\label{sect:models}
In addition to giving an independent constraint on the eccentricity 
via the phase of the secondary eclipse, the depth of the eclipse is 
also a useful measure of the atmospheric properties of the companion. 
In particular, once the temperature of the companion is known, e.g., 
from other eclipse measurements in the infrared, the geometric albedo 
$A_{\rm g}$ can be estimated. The total eclipse depth can be represented as 
follows \citep[see, e.g.,][]{cowan2011, daylan2021} 
%
%
\begin{eqnarray}
\label{eq:tot_depth}
\delta_{\rm obs} = \delta_{\rm refl} + \delta_{\rm therm}
\hspace{2mm}.  
\end{eqnarray}
Because of the internal energy source in the case of BDs, the second 
term (that includes also the absorbed stellar flux) yields an important 
contribution to the observed eclipse depth -- in particular, in the 
infrared wavebands. 
Considering that the eccentricity for these BDs are not negligible, 
the eclipse depth, associated with the reflected light, is related 
to the geometric albedo $A_{\rm g}$ through the following relations 
%
%
\begin{eqnarray}
\label{eq:galbedo}
\delta_{\rm refl} = A_{\rm g} \left({R_{\rm c} \over r_{\rm occ}}\right)^2 \hspace{2mm}, \hspace{2mm} 
r_{\rm occ}={a(1-e^2) \over 1-e\sin\omega} \hspace{2mm},   
\end{eqnarray}
where $R_{\rm c}$ is the radius of the companion (i.e., that of the BD), $a$ 
is the semi-major axis, $e$ is the eccentricity, $\omega$ is the 
argument of the periapsis and $r_{\rm occ}$ is the distance of the companion 
from the star at the moment of the occultation. Although not shown 
explicitly, $A_{\rm g}$ is wavelength-dependent, leading to effective 
reflectivity at short wavelengths, and rather ineffective reflectivity 
further down in the infrared regime. We approximately take this dependence 
into consideration by using the models employed on the occultation 
data of HD~189733b by \cite{krenn2023}. Their Fig.~6 is used to 
estimate the efficiency of the reflection of the incident stellar 
flux as a function of wavelength and overall (gray) $A_{\rm g}$ values. 
Their result implies that the amount of the reflected flux is less 
than $\sim 20$\% in the Spitzer wavebands than that of in the Kepler 
waveband even if the gray geometric albedo is as high as $75$\% in 
the Kepler waveband. 

The secondary eclipse depth due to the re-radiation part of the stellar 
flux and the internal energy source is given by 
%
%
\begin{eqnarray}
\label{eq:t_int}
\delta_{\rm therm} = \left({R_{\rm c} \over R_{\rm s}}\right)^2 {F_{\rm c}(\lambda,T_{\rm day}) \over F_{\rm s}(\lambda)}
\hspace{2mm}.   
\end{eqnarray}
Here $R_{\rm s}$ is the stellar radius, $F_{\rm c}$ and $F_{\rm s}$ are the fluxes 
of the companion and the star respectively and $\lambda$ is the wavelength.  
The day-side companion temperature $T_{\rm day}$ depends on the temperature 
generated by the internal heat source $T_{\rm int}$ and on $T_{\rm irr}$, the 
temperature originating from the non-reflected part of the stellar 
irradiation. We relate these temperatures following, e.g., \cite{amaro2023}
%
%
\begin{eqnarray}
\label{eq:day_depth}
T_{\rm day}^4 = T_{\rm irr}^4 + T_{\rm int}^4 \hspace{2mm}.   
\end{eqnarray}
To compute $T_{\rm irr}$, we follow the customary parameterization \citep[i.e.,][]{cowan2011}  
%
%
\begin{eqnarray}
\label{eq:irr_temp}
T_{\rm irr}^4 = \alpha T_0^4 \hspace{2mm}, \hspace{2mm} \alpha=(1-A_{\rm B})\left({2 \over 3}-\epsilon {5 \over 12}\right)  
\hspace{2mm},    
\end{eqnarray}
where the substellar temperature $T_0$ due to stellar irradiation can be 
evaluated with the knowledge of the stellar effective temperature $T_{\rm s}$ 
%
%
\begin{eqnarray}
\label{eq:day_temp}
T_0 = T_{\rm s}\sqrt{R_{\rm c}/r_{\rm occ}} \hspace{2mm}.   
\end{eqnarray}
The absorbed heat is controlled by the Bond albedo $A_{\rm B}$, whereas the 
efficiency of the day/night circulation is determined by the parameter 
$\epsilon$. Although the effect of the stellar irradiation on 
$T_{\rm day}$ is under $4$\% for both objects studied in this paper, we 
use $T_{\rm day}$ as given by Eq.~\ref{eq:day_depth} for the effective 
temperature of the BD atmosphere models, partially accounting for the 
effect of stellar irradiation on the theoretical spectra. 

The Bond albedo is a highly involved quantity 
\citep[e.g.,][]{marley1999} and it is hard to relate to the geometric 
albedo in a precise way. Nevertheless, the two types of albedos are 
obviously related. To take this relation into consideration in a 
rather approximate way, we considered two classes of objects. First 
we examined $15$ solar system bodies as collected at the corresponding 
Wikipedia site\footnote{\url{https://en.wikipedia.org/wiki/Bond_albedo}} 
\citep[see also][for the planets]{cox2000}. Omitting Haumea and Enceladus, 
we found $A_{\rm g} \sim 1.1 A_{\rm B}$, with a substantial standard deviation of 
$\sim 0.1$. Then we took the hot Jupiter sample of \cite{schwartz2015} 
(their Table~4, with fully corrected $A_{\rm g}$ and assumed gray planet 
atmosphere (columns five and and eight, respectively). Here we found a 
very tight correlation for the $11$ planets, but with a lower slope of 
$\sim 0.8$. Considering that this estimate is based on relatively early 
sets of hot Jupiter atmospheric observations, and that the solar system 
albedos can be regarded as less relevant for BDs, we think that the 
$A_{\rm g}=A_{\rm B}$ approximation in the Kepler waveband can be a reasonable one 
in the context of this paper.  
In any case, the specific relation of $A_{\rm g}$ and $A_{\rm B}$ has only a 
moderate effect on the final conclusion, due to the high value of 
$T_{\rm int}$ and the high power used for the temperature components in 
calculating the planet temperature (see Eq.~\ref{eq:day_depth}). 

To utilize the above formulae in the estimation of the theoretical 
eclipse depths, we need stellar and sub-stellar atmosphere models 
yielding the corresponding spectra as they appear in 
Eq.~\ref{eq:t_int}. We note that the simple black body fluxes 
perform rather poorly, because of the significant molecular bands  
in the sub-stellar, and, for similar reason, in the low stellar 
mass regimes. For the stellar spectra we used the Kurucz2003 
\citep[][]{castelli2003} and BT-NextGen models by \cite{allard2012}. 
For the two BD systems and for the wavelength regime used, they 
yield very similar results. For the companions we took the quite 
recent ATMO models by \cite{phillips2020}. These are ``clear'' 
(i.e., non-cloudy), non-irradiated models with chemical 
equilibrium and solar composition. For a brief justification 
of our choice of cloudless models without stellar irradiation we 
note the following.  

\cite{beatty2018} used irradiated cloudy and cloudless models for 
EPIC~219388192. In the left panel of their Fig.~8 the cloudless, 
but irradiated models reach a maximum eclipse depth of $\sim 1600$~ppm 
between the two Spitzer passbands. Using their preferred internal 
temperature of $1700$~K with the same atmospheric conditions 
(low $A_{\rm B}$ of $1$\%, no day/night heat redistribution), we get an 
eclipse depth of $1500$~ppm. One can get an exact agreement with 
their model value by increasing $T_{\rm int}$ to $1800$~K, but we think 
that this is well within the subtle differences of the theoretical 
models and do not influence our broad estimate of the geometric 
albedo of the BD component, the focus point of this work.  
 
For using cloudless models, we again refer to Fig.~8 of 
\cite{beatty2018}. Comparing the left (``clear'') and right (``cloudy'') 
spectra, we see that there is a $\sim 10$\% reduction in the emitted 
flux for the cloudy models. Similarly, we infer from a brief inspection 
of Fig.~10 from \cite{morley2024} that the effect of clouds near to 
the optical part of the emission spectrum may also not exceed $\sim10$~\%.  

Importantly, there are also evolution models that yield theoretical 
values for $T_{\rm int}$, directly related to the observed secondary 
eclipse depths. Because this estimate comes from the independently 
measured companion mass, radius, age and assumed chemical composition, 
the derived $T_{\rm int}$ value is very important in harmonizing 
evolutionary and atmosphere models with the observations. 
To predict $T_{\rm int}$ from models, we used the very recent SONORA 
models by \cite{davis2025} \citep[see also][]{marley2021}. For 
comparison, we also utilized the earlier models of \cite{baraffe2003}. 
We used linear and quadratic interpolations both for the atmosphere 
and for the evolutionary models to match the input parameters exactly. 
Because of the deeper than expected eclipse depths, for $T_{\rm irr}$ 
we assumed that the day-to-night heat transport is ineffective, i.e., 
$\epsilon=0$. This is in agreement with the approach of \cite{beatty2018} 
and also supported by the low efficiency of the energy transport 
suggested by the secondary eclipse observations at $2\mu m$ of Hot 
Jupiters \citep{kovacs2019}. 

In the following subsections we examine the constraints put on 
the geometric albedos by the above relations. In this comparison 
we do not intend to perform a full-scale modeling (e.g., comparing 
various atmospheric compositions), because: firstly, it is 
out of the scope of the paper, and, secondly, and even more 
importantly, the limited data (too few spectral points with low 
accuracy in relatively wide filter bands) would not justify such 
a deep study.    
%
%
\subsection{The high geometric albedo of EPIC 219388192b}
\label{sect:high_albedo}
Together with the Spitzer observations, we have three secondary 
eclipse measurements for this object: with depths $1147\pm 213$ ppm  
(at $3.5\mu m$) and $1097\pm 225$ ppm at $4.4\mu m$\footnote{We note 
that \cite{beatty2018} use mean wavelength, whereas we use effective 
wavelength; the latter are shorter by $0.1\mu m$.} as given by 
\cite{beatty2018} and $70\pm 12$ ppm at $0.6\mu m$ from the analysis 
presented in this paper. 
%
%
\begin{figure}[t]
\centering
\includegraphics[width=0.49\textwidth]{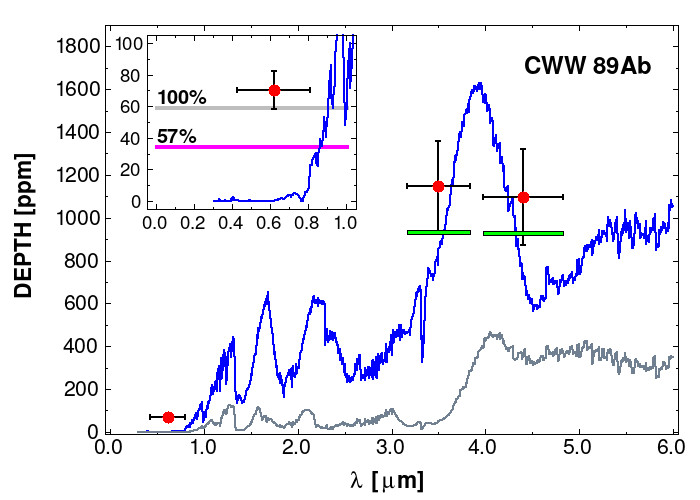}
\caption{Theoretical emission spectra transformed to eclipse depth 
         for EPIC~219388192b (CWW~89Ab). Only the thermal 
	 components are included (internal and absorbed stellar 
	 sources). Blue: $T_{\rm int}=1900$~K -- fit to the Spitzer 
	 data within $1\sigma$; gray: $T_{\rm int}=944$~K -- BD evolutionary 
	 temperature. For the star we used the BT-NextGen models 
	 \citep{allard2012}, whereas for the BD we employed the 
	 ATMO models of \cite{phillips2020}. Green horizontal bars show 
	 the theoretical depths in the Spitzer wavebands. Red dots 
	 with error bars show the observed depths by \cite{beatty2018}. 
	 The third red dot in the left bottom corner is the eclipse 
	 depth in the Kp filter. The inset zooms in on this region, 
	 indicating the expected eclipse depth for the reflectivity 
	 levels shown. Horizontal error bars indicate the equivalent 
	 filter width. See text for further details.} 
\label{em_spec_219388192}
\end{figure}
Based on its cluster membership, the age is reasonably fixed at 
$2.67\pm 0.47$~Gyr \citep[i.e., ][]{torres2021}. From the system 
analysis, \cite{carmichael2023} obtained $39.2\pm 1.1$~M$_{\rm J}$ for the 
mass of the BD component. By using these values, from the SONORA 
models we get $R_{\rm BD}=0.91$~R$_{\rm J}$, in excellent agreement with 
the observed value of $0.94\pm 0.02$ given by \cite{carmichael2023}. 
From the same models, for $T_{\rm int}$ we get $944$~K. Using this 
$T_{\rm int}$ and assuming $A_{\rm B}=0.6$, from the ATMO models we get the 
spectrum shown in Fig.~\ref{em_spec_219388192} by a gray line. 
We see, that with the evolutionary internal temperature, 
the measured Spitzer fluxes are several sigmas apart 
from the fluxes predicted by the BD atmosphere models. Because of 
the relatively low value of $T_{\rm int}$ and the proximity of the 
host star, it matters how the Bond albedo is 
chosen.\footnote{We recall that, meanwhile, the reflective component 
at the Spitzer passband remains small -- less than $10$~ppm.}
 
In the example shown in Fig.~\ref{em_spec_219388192}, the Bond albedo 
was chosen to yield the $3\sigma$ low limit of the observed eclipse 
depth in the Kp band, and, assuming that the $A_{\rm B}=A_{\rm g}$ condition holds. 
Allowing $A_{\rm B}$ varying freely between $0.1$ and $0.9$, the corresponding 
predicted eclipse depths vary in $225$--$59$~ppm (at $3.5\mu m$) 
and in $522$--$267$~ppm (at $4.4\mu m$). In both extremes the 
theoretical fluxes significantly underestimate the observed values, 
especially in the shorter waveband. These results support the 
observation of \cite{beatty2018} that EPIC 219388192b shows 
serious overluminosity with respect to the evolutionary models. 

By tuning upward $T_{\rm int}$, we can match the observed Spitzer fluxes 
quite closely. By scaling the peak flux to the same value as in the 
models of \cite{beatty2018}, we find that for the ATMO models we 
need to set $T_{\rm int}$ by $200$~K higher than the one preferred by 
\cite{beatty2018}. With $T_{\rm int}=1900$~K, we can fit the Spitzer 
fluxes within $1\sigma$. Of course, depending on the accepted 
accuracy within which the fit is considered to be acceptable, 
a lower $T_{\rm int}$ value is also possible, but from the current 
Spitzer data the disagreement between the evolutionary and observed 
internal temperatures seems to be quite solid. 

As seen in Fig.~\ref{em_spec_219388192}, the theoretical thermal 
flux has negligible contribution at the Kepler waveband even at the 
high temperature required by the Spitzer data. It follows that the 
only way to produce the observed eclipse depth, is if the geometric 
albedo is high. The inset in Fig.~\ref{em_spec_219388192} shows the 
relation between the observed and theoretically expected eclipse 
depths. At $A_{\rm g}=1.0$, the expected eclipse depth is $59$~ppm. 
In the extreme case, if the depth is overestimated by $3\sigma$, 
i.e., not $70$~ppm, but only $34$~ppm, the resulting $A_{\rm g}$ is 
$0.57$, which is still remarkably high. It is worth noting 
that KELT-1b may have also a similarly large reflectivity in the 
TESS band according to \cite{beatty2020}, although this has been 
challenged recently by the reanalysis of the TESS data by 
\cite{essen2021} and the inclusion of the CHEOPS observations 
by \cite{parviainen2022}. Some of these discrepancies may be 
attributed to seasonal variations in the reflective clouds of 
KELT-1b \citep[i.e.,][]{parviainen2023}.     
%
%
%
\subsection{The unconstrained albedo of EPIC 211946007b}
\label{sect:normal_albedo}
Unfortunately, for this system, the only secondary eclipse data 
available to us are those presented in this paper. Because the  
single, short-waveband eclipse depth is insufficient to derive 
$T_{\rm int}$, we must rely on the evolutionary models. Since this 
target is also a cluster member we can follow the same steps as 
in the case of EPIC 219388192. Being member of the Praesepe, the 
age of the system is $0.64\pm 0.02$~Gyr \citep[][]{morales2022}. 
From the system analysis \citep{carmichael2023} we have 
$54.6\pm 6.8$~M$_{\rm J}$ for the mass of the BD component. From these 
data the SONORA models yield $R_{\rm BD}=1.01$~R$_{\rm J}$, within $1\sigma$ 
of the observed value of $0.95\pm 0.07$~R$_{\rm J}$ \citep{carmichael2023}. 
The match also yields $T_{\rm int}=1624$~K. 

%
%
\begin{figure}[t]
\centering
\includegraphics[width=0.45\textwidth]{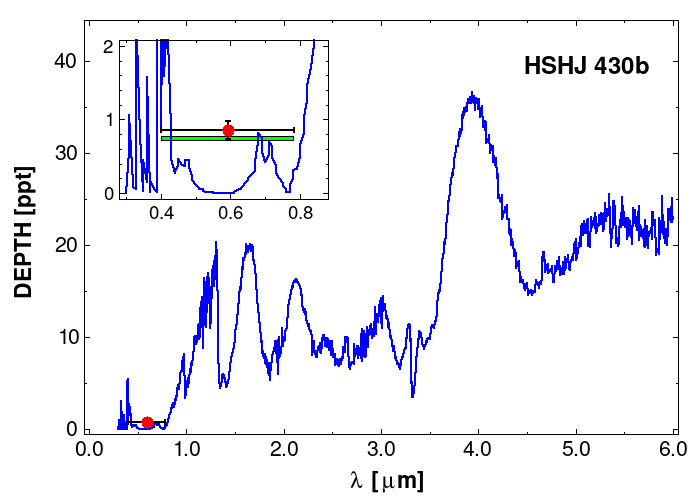}
\caption{As in Fig.~\ref{em_spec_219388192} (same theoretical models 
         and figure setting), but for EPIC 211946007b (HSHJ~430b). 
	 We used $100$~K higher internal temperature (i.e., $T_{\rm int}=1724$~K) 
	 than the theoretical evolutionary value. The green  
	 horizontal bar in the inset shows the theoretical depth 
	 in the Kepler waveband, assuming $10$~\% Bond albedo. 
	 Red dots with error bars show the observed depth.} 
\label{em_spec_211946007}
\end{figure}

The observed eclipse depth in the Kp band is $852\pm 123$~ppm. 
With $T_{\rm int}$ as above and assuming ``standard'' low $A_{\rm g}$ of 
$0.1$, and apply the $A_{\rm B}=A_{\rm g}$ assumption, we get a depth of 
$511$~ppm, which is within $3\sigma$ of the observed value. 
We need to increase $A_{\rm g}$ to $0.6$ to come within the $1\sigma$ 
neighborhood of the observed value. Alternatively, we may assume 
that $T_{\rm int}$ is slightly underestimated by the models 
\citep[indeed,][ yields $200$~K higher value]{baraffe2003}. 
Allowing an increase of only $100$~K in $T_{\rm int}$ and setting 
$A_{\rm g}=0.2$, we arrive at $797$~ppm, well within the $1\sigma$ 
limit of the observed eclipse depth. If we assume complete 
heat redistribution (i.e. $\epsilon=1$, instead of $0$, as we 
assumed throughout the paper), then, with the same $A_{\rm g}$, we 
get a depth of $775$~ppm, again in good agreement with the 
observed depth. We conclude that the available data for 
EPIC 211946007b are not in contradiction with the low albedo, 
standard evolutionary internal temperature scenario. 

For completeness, in Fig.~\ref{em_spec_211946007} we show the 
emission spectrum of EPIC 211946007 for $T_{\rm int}=1724$~K and 
$A_{\rm g}=0.1$ together with the observed eclipse depth as given in 
this paper. For the first sight it may look inconsistent that 
the integrated theoretical emission in the neighborhood of 
$0.6$~$\mu m$ can indeed reach the value of the observed flux. 
However, a closer examination of the Kp filter function shows 
that although the transmission decreases considerably outside 
the effective waveband (shown by the horizontal bars in the 
figure), the emission sharply increases in these regions and 
this compensates for the decrease in the filter transmission. 
The comparison of Figures~\ref{em_spec_219388192} and 
\ref{em_spec_211946007} shows that while topologically the 
two spectra are very similar, because of the substantially 
younger age of EPIC 211946007b, its overall thermal emission 
is much higher than that of EPIC 219388192b. This difference 
allows the low albedo scenario quite more likely for 
EPIC 211946007b.  
%
%
%
\section{Conclusions}
\label{sect:conclude}
The earlier perception on the low occurrence rate of brown dwarf 
companions in short-period binaries \citep{marcy2000, halbwachs2000}  
seems to be considerably modified by the increasing number of 
discoveries during the past decade. The contribution of the 
discoveries via the TESS satellite is especially spectacular. 
More than half of the currently known $\sim 55$ BDs in 
eclipsing binary sytems have been discovered by transit 
searches on the photometric time series based on the 
observations made by TESS. The followup observations generally 
yield orbital elements and masses, but secondary eclipse detections 
are relatively rare, in spite of the expected higher signal-to-noise 
ratio (with respect to the thermal emission of planets). Secondary 
eclipse observations are instrumental in the verification of 
atmospheric and evolutionary model predictions. 

We used multiple datasets to detect the secondary eclipses of two 
binaries hosting BDs. Both systems were observed by K2 (the Kepler 
two-wheel mission) and reside in well-known open clusters. In 
addition, there are several independent photometric datasets 
available for both systems. Furthermore, EPIC 211946007 (HSHJ 430) 
was observed during three different K2 campaigns. 

Once the variabilities due to instrumental systematics, stellar 
spots and transits were filtered out, the data were searched 
for secondary eclipses on data source and campaign field bases. 
Then, the separate secondary eclipse search statistics were merged 
to increase the signal-to-noise ratio of the detected signal. To 
verify the consistency of the method and to obtain another estimate 
of the eclipse signal, we performed similar data merging for the 
light curves. In these two approaches the averaging relies on different 
pieces of information of the constituent datasets. Therefore, checking 
the consistency of the results is a significant step in confirming 
the reliability of the detections. The main results of the paper 
can be summarized as follows. 
  
\begin{itemize}
\item[$-$]
EPIC 219388192 (CWW 89A) shows a secondary eclipse with a depth of 
$(70\pm 12)$~ppm. The thermal emission \citep[even at the excessive 
internal temperature predicted from the Spitzer observations by][]{beatty2018} 
is not enough in the Kp band to explain this depth. Therefore, it 
seems unavoidable to conclude that this BD is also peculiar in that 
it has a high geometric albedo. The minimum albedo allowed by the 
eclipse depth is 0.6. Obviously, independent measurements at short 
wavebands would be quite important to confirm or refute our finding. 
\item[$-$]
EPIC 211946007 (HSHJ 430) did not have a secondary eclipse 
observation before this report. We detected a depth of 
$(852\pm 123)$~ppm which, due to the young age of the system, 
can be produced (almost entirely) by internal heat. Therefore, 
the `standard' low-albedo scenario is quite likely for the BD 
component of this system. 
\end{itemize}
%

%
%
\begin{acknowledgements}
%
We thank the referee for the thorough and instructive report 
which contributed to the improvement of the clarity of the paper.   
%
This work has been inspired by the TESS III Science Conference held 
at MIT in July, 2024. Discussions with Theron Carmichael, Dave Latham, 
Avi Shporer, Jon Jenkins and Eric Feigelson were very stimulating. 

The help given by Melanie A. Swain regarding the usage of the current 
ExoFOP site is greatly appreciated. 

The availability of independently obtained photometric time series was 
essential in the successful detections presented in this paper. Thanks 
are due to Eric Petigura, Andrew Vanderburg and Rodrigo Luger for making 
their data accessible to the public.   

This research has made use of the NASA/IPAC Infrared Science Archive, 
which is funded by the National Aeronautics and Space Administration 
and operated by the California Institute of Technology.

This research has made use of the Spanish Virtual Observatory 
(\url{https://svo.cab.inta-csic.es}) project funded by 
MCIN/AEI/10.13039/501100011033/ through grant PID2020-112949GB-I00.

This paper includes data collected by the Kepler mission and obtained 
from the MAST data archive at the Space Telescope Science Institute 
(STScI). Funding for the Kepler mission is provided by the NASA 
Science Mission Directorate. STScI is operated by the Association 
of Universities for Research in Astronomy, Inc., under NASA contract 
NAS 5-201326555.
\end{acknowledgements}

%
%

%

%
%
%
\begin{appendix}
\section{Average flux correction}
\label{app_A}
Due to the variation of the background star contamination because of the 
different aperture masks, the different data sources yield different 
transit depths. We considered this effect at the stage of data input. 
For a given data source, object and K2 campaign number, we applied the 
following type of correction to the input SAP fluxes $FLX$: 
\begin{eqnarray}
\label{eq:f_flux}
FLX_c = FLX\times f_c/<FLX> \hspace{2mm},    
\end{eqnarray}
where $<FLX>$ is the robust average of the input flux and $f_c$ is the 
correction factor. This factor is set by a trial and error approximation 
by inspecting the final data product -- resulting from the various noise 
and signal filtering steps (see Sect.~\ref{sect:method1}). In Table~\ref{FC} 
we summarize the set-by-set values of $f_c$. The factors were chosen so 
that the respective transit depths be the same for the given target, 
independently of the data source. We used the following transit depths 
for the two targets. For EPIC 219388192: $\delta_1=0.0104$ and for 
EPIC 211946007: $\delta_1=0.1400$. 
%
%
\begin{table}[h]
\begin{flushleft}
\caption{Input flux correction factors}
\label{FC}
\scalebox{1.0}{
\begin{tabular}{cccccc}
\hline
Field & Source &  $f_c$ & Field & Source & $f_c$ \\
\hline\hline
\multicolumn{3}{c}{EPIC 211946007} & \multicolumn{3}{c}{EPIC 219388192}\\
\hline
c05 &  PDC  &  0.980 & c07 &  PDC  &  1.000\\
c05 &  PET  &  1.042 & c07 &  PET  &  1.085\\
c05 &  VAN  &  1.027 & c07 &  VAN  &  1.138\\ 
c05 &  LUG  &  0.996 & c07 &  LUG  &  1.000\\
\hline
c16 &  PDC  &  0.975\\ 
c16 &  PET  &  0.975\\
c16 &  VAN  &  0.985\\ 
c16 &  LUG  &  0.997\\
\hline  
c18 &  PDC  &  0.985\\
c18 &  PET  &  1.000\\
c18 &  VAN  &  1.008\\ 
c18 &  LUG  &  1.057\\
\hline
\end{tabular}}
\vskip 4 pt
\end{flushleft}
\end{table}  
\end{appendix}
\end{document}